\documentclass[conference]{IEEEtran}
\IEEEoverridecommandlockouts
\usepackage{cite}
\usepackage{amsmath,amssymb,amsfonts}
\usepackage{algorithmic}
\usepackage{graphicx}
\usepackage{textcomp}
\usepackage{xcolor}

\usepackage{tabularx}
\usepackage{adjustbox}
\usepackage{cleveref}
\usepackage{fancyvrb}

\usepackage{booktabs}
\usepackage{multirow}
\usepackage{float}
\usepackage{svg}
\usepackage{enumitem}

\def\BibTeX{{\rm B\kern-.05em{\sc i\kern-.025em b}\kern-.08em
    T\kern-.1667em\lower.7ex\hbox{E}\kern-.125emX}}

\usepackage{fancyhdr,lipsum}
\fancypagestyle{firstpage}{% Page style for first page
  \fancyhf{}% Clear header/footer
  \fancyhead[C]{To appear at the 3rd International Conference on Emergent Quantum Technologies (ICEQT'24), July 2024.}% Header
  \fancyfoot[C]{\thepage}% Footer
}

\begin{document}

\title{A Comparative Analysis of Hybrid-Quantum Classical Neural Networks}

%\author{Anonymous Authors}

% \author{Kamila Zaman\textsuperscript{1,2,*}, 
% Tasnim Ahmed\textsuperscript{1,2,*},
% Muhammad Abdullah Hanif\textsuperscript{1,2}, 
% Alberto Marchisio\textsuperscript{1,2}, 
% and Muhammad Shafique\textsuperscript{1,2}, 
% }
% \affiliation{%
%   \institution{\textsuperscript{1}eBrain Lab, Division of Engineering, New York University Abu Dhabi (NYUAD), Abu Dhabi, UAE}
%   %\institution{\textsuperscript{2}Center for Quantum and Topological Systems (CQTS), NYUAD Research Institute, New York University Abu Dhabi, PO Box 129188, Abu Dhabi, UAE}
%   %\city{New York}
%   \country{\textsuperscript{2}Center for Quantum and Topological Systems (CQTS), NYUAD Research Institute, NYUAD, Abu Dhabi, UAE}
%   }
% \email{{kz2137, tasnim.ahmed, mh6117, alberto.marchisio, muhammad.shafique}@nyu.edu}

\author{\IEEEauthorblockN{Kamila Zaman\textsuperscript{1,2,*}, 
Tasnim Ahmed\textsuperscript{1,2,*},
Muhammad Abdullah Hanif\textsuperscript{1,2}, 
Alberto Marchisio\textsuperscript{1,2}, \\
and Muhammad Shafique\textsuperscript{1,2}}
\IEEEauthorblockA{\textsuperscript{1}eBrain Lab, Division of Engineering, New York University Abu Dhabi (NYUAD), Abu Dhabi, UAE \\
\textsuperscript{2}Center for Quantum and Topological Systems (CQTS), NYUAD Research Institute, NYUAD, Abu Dhabi, UAE\\
email: \{kz2137, tasnim.ahmed, mh6117, alberto.marchisio, muhammad.shafique\}@nyu.edu
}
\thanks{*These authors contributed equally to this work.}
% \and
% \and
% \and
% \and
% \and
}

\maketitle
\thispagestyle{firstpage}

\begin{abstract}
Hybrid Quantum-Classical Machine Learning (ML) is an emerging field, amalgamating the strengths of both classical neural networks and quantum variational circuits on the current noisy intermediate-scale quantum devices~\cite{zaman2023survey}. This paper performs an extensive comparative analysis between different hybrid quantum-classical machine learning algorithms, namely Quantum Convolution Neural Network, Quanvolutional Neural Network and Quantum ResNet, for image classification. % over two different quantum computing frameworks, i.e., Qiskit \& PennyLane. 
The experiments designed in this paper focus on different Quantum ML (QML) algorithms to better understand the accuracy variation across the different quantum architectures by implementing interchangeable quantum circuit layers, varying the repetition of such layers and their efficient placement. Such variations enable us to compare the accuracy across different architectural permutations of a given hybrid QML algorithm. %The accuracy comparison of the hybrid models is based on the accuracy, loss and execution time of the training process, which 
The performance comparison of the hybrid models, based on the accuracy, 
provides us with an understanding of hybrid quantum-classical convergence in correlation with the quantum layer count and the qubit count variations in the circuit.
\end{abstract}


\section{Introduction}

Merging Quantum Computing (QC) with Machine Learning (ML) forms the emerging Quantum Machine Learning (QML) paradigm~\cite{Schuld_2019, Schetakis_2022SciRep_ReviewQML, Markidis_2023Entropy_ProgrammingQNNs, Massoli_2023CSUR_LeapQCQNN}. It represents an excellent opportunity for researchers and industries to make phenomenal discoveries and unravel efficient ways to solve complex real-world problems with significant direction towards practicality and improved accuracy and robustness compared to classical systems~\cite{maouaki2024advqunn}. QML opens new avenues for the community to discover, build, and align their designs to different levels of the quantum stack. However, the currently developed Noisy-Intermediate Scale Quantum (NISQ) devices~\cite{Preskill_2018arxiv_QC_NISQ_era} have a limited number of qubits, with small-scale resilience to noise, therefore, making it difficult to develop and practically realize the potential of standalone QML algorithms. The limitation of NISQ devices has motivated the development of hybrid quantum-classical machine learning algorithms (HQML)~\cite{zaman2023survey}, which are NISQ-compatible algorithms.

\begin{figure}[!t]
\centering
\includegraphics[width =\linewidth]{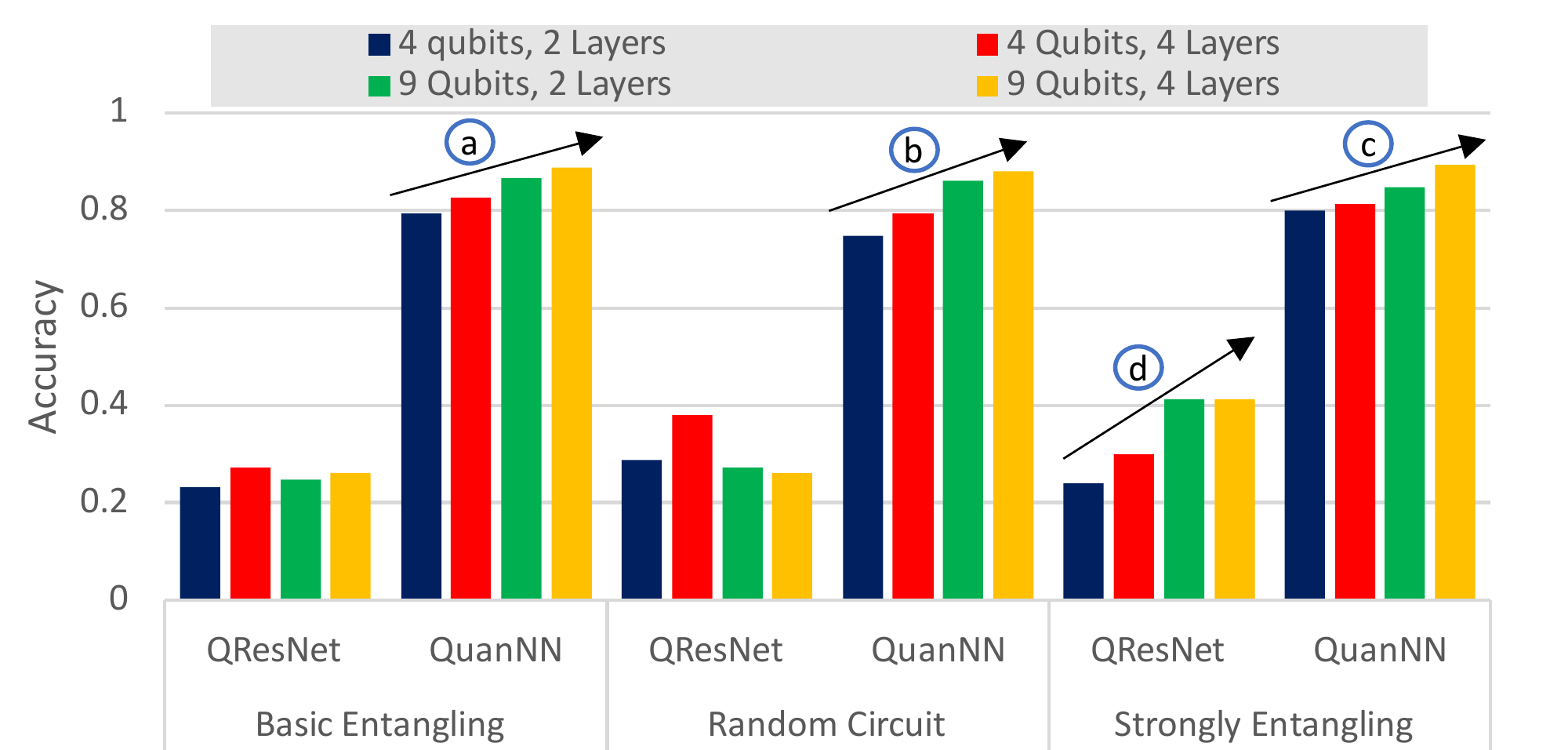}
\caption{\textbf{Penny-lane:} The arrows indicate an observed pattern suggesting a potential positive correlation of accuracy with the number of layers \& qubits in hybrid quantum-classical architectures. Specifically, (a), (b), and (c) consistently exhibit this behavior for the QuanNN across all entangling circuits, while (d) displays a similar effect in the QResNet. While the QuanNN responds to both layer and qubit count, the QResNet demonstrates improved accuracy primarily with an increase in qubits only. Hence, observation at (d), raises questions for further studies on whether the QResNet's accuracy significantly improves with a substantial increase in qubits further or not. }
\label{fig:PennyLane_motivation}
\end{figure}

HQML algorithms have emerged as an important paradigm that amalgamates the power of both classical and quantum processing for machine learning tasks, opening new avenues for algorithm and architecture exploration~\cite{bergholm2022pennylane}. The most renowned HQML algorithms employ variational quantum circuits, featuring parameterized quantum gates optimized by classical computers to achieve specific goals. These variational quantum circuits in QML, known as Quantum Neural Networks (QNNs), hold significant promise due to their expressiveness and reduced trainable parameters, garnering considerable development interest~\cite{zaman2024studying, innan2024fedqnn}.

Given the recent breakthroughs of classical algorithms such as Convolutional Neural Networks (CNNs) for image classification tasks, numerous QNN architectures have been developed for classification tasks. However, those architectures are either implemented as basic building blocks or lack the verification of their usability against a benchmark model for classification tasks. \textit{The observations in Fig.~\ref{fig:PennyLane_motivation}, based on our implementation of permutations of two different algorithms, show that different architectures of the same algorithms have different impacts on classification tasks.} Thus, it is imperative to explore the architectures, applications, and usefulness of QNN algorithms and evaluate their accuracy to identify efficient model configurations that can be deemed as reference benchmarks for future research. 

In this paper, we explore three different QNN algorithms amongst the pool of hybrid algorithms, namely, Quanvolutional Neural Networks~\cite{henderson2019quanvolutional}, Quantum Convolutional Neural Networks~\cite{Cong_2019}, and Quantum ResNet~\cite{liang2021hybrid} and perform an extensive comparative analysis. Our methodology first involves identifying efficient architectures amongst the commonly used QNNs and understanding their practical utility for classification tasks. Then we assess how variations of such algorithms impact their accuracy and robustness under architectural permutations of each algorithm. The architectural permutations are based on the implementation of interchangeable variational circuit layers over different qubit counts, varying repetition of the layers, and considering their optimal placement. By implementing the varied models, we evaluate the performance of QNNs based on the accuracy %, loss, and execution time 
of the training process, which provides us with an understanding of hybrid quantum-classical convergence in correlation with our experimental approach. Such a comprehensive analysis is necessary to establish an understanding of the correlation between circuit architectures, their robustness, and utility in QML.

\begin{figure}[!t]
    \centering
    \includegraphics[width=\linewidth]{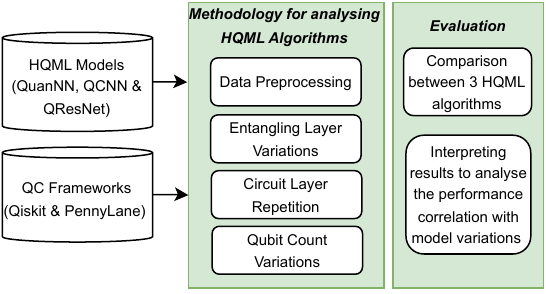}
    \caption{Overview of our novel contributions.}
    \label{fig: novel_contributions}
\end{figure}

\subsection{Our Novel Contributions}

An overview of our novel contributions is shown in Fig.~\ref{fig: novel_contributions}. Their brief descriptions with key features is presented below. 

\begin{itemize}[leftmargin=*]
    \item We propose a methodology to investigate QNN models' circuit permutations and their effect on the accuracy for classification tasks. (\textbf{Section \ref{sec:methodology}})
    \item We analyze the effect of different circuit layer variations\footnote{The circuit variations are studied only for QNN models that support different types of entanglement.}, namely, Random Circuit, Basic Entangling, and Strongly Entangling. (\textbf{Section \ref{sec:entangling_eircuit_variations}})
    %\item We design a methodology to investigate entangled circuit layer variations, namely, random entangling, basic entangling \& Strongly Entangling, however only for algorithms that allow different kinds of entanglement. (\textbf{Section \ref{sec:entangling_eircuit_variations}}) This allows us to analyze the effect of different entangling methods on the classification task. (\textbf{Section \ref{sec:entangling_eircuit_variations}})
    \item We investigate the repetition of different layers to analyze the effect of circuit depth and complexity. (\textbf{Section \ref{sec:layer_count_variations}})
    \item We test the scalability of the algorithms by varying qubit counts (\textbf{Section \ref{sec:qubit_count_variations}}) 
    
\end{itemize}

\section{Selected Hybrid QML Algorithms}
\label{sec:algorithms}

\begin{figure}[h]
    \centering
    \includegraphics[width=\linewidth]{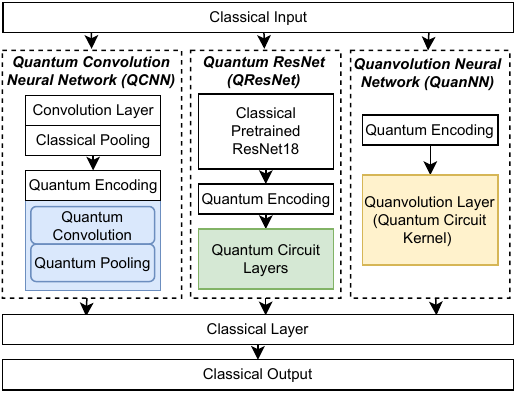}
    \caption{Pipeline of the implemented hybrid QML algorithms. Each model utilizes a classical fully connected layer to transform quantum circuit measurement into classification probabilities. In the QCNN, classical convolutional and pooling layers are used for image downsizing to match the qubit count of a circuit.}
    \label{fig:hybrid_qml_algorithms}
\end{figure}

An overview of the hybrid QML algorithms is shown in Fig.~\ref{fig:hybrid_qml_algorithms}. A brief description of the key features of each algorithm and their implementations is presented in the following paragraphs.
%\textit{Note: We implemented all these pipelines in Qiskit~\cite{Qiskit} \& PennyLane~\cite{bergholm2022pennylane} to enable the support for both QML frameworks.}

\subsection{Quanvolutional Neural Networks}%~\cite{henderson2019quanvolutional}}
The Quanvolutional Neural Network (QuanNN) is an innovative hybrid quantum-classical architecture developed in~\cite{henderson2019quanvolutional}, which enhances the capabilities of classical CNNs by harnessing the potential of quantum computation. This architecture introduces a new type of transformation layer called the quanvolutional layer, akin to classical convolutional layers, composed of multiple quanvolutional filters which locally transform input data, extracting valuable features for classification. These filters correspond to a certain circuit design, which can either be generated randomly or based on a specific entanglement, namely Basic Entangling or Strongly Entangling. The reason we chose the QuanNN as one of our benchmarking models is because of its generalizability which can be achieved by adhering to the following conditions: specifying an arbitrary integer number of quanvolutional filters in each layer, stacking multiple quanvolutional layers in the network, defining layer-specific parameters like encoding method, entanglement, and average quantum gates per qubit in the quantum circuit. 
To formalize classical data transformation using quanvolutional filters: 
\begin{enumerate}[leftmargin=*]
    \item Start with a single filter \textit{q} operating on subsections \textit{ux} of dataset images.
    \item Encode \textit{ux} into an initialized state \textit{ix} using an encoding function~\textit{e}.
    \item Apply the quantum circuit to \textit{ix}, producing an output quantum state \textit{ox}.
    \item Decode \textit{ox} to ensure consistent outputs, resulting in the final decoded state \textit{fx}.
    \item  This entire process, denoted as the ``quanvolutional filter transformation'' \textit{Q}, is \mbox{\textit{fx = Q(ux, e, q, d)}}.
    
\end{enumerate}

\begin{figure*}[h]
    \centering
    \includegraphics[width=\linewidth]{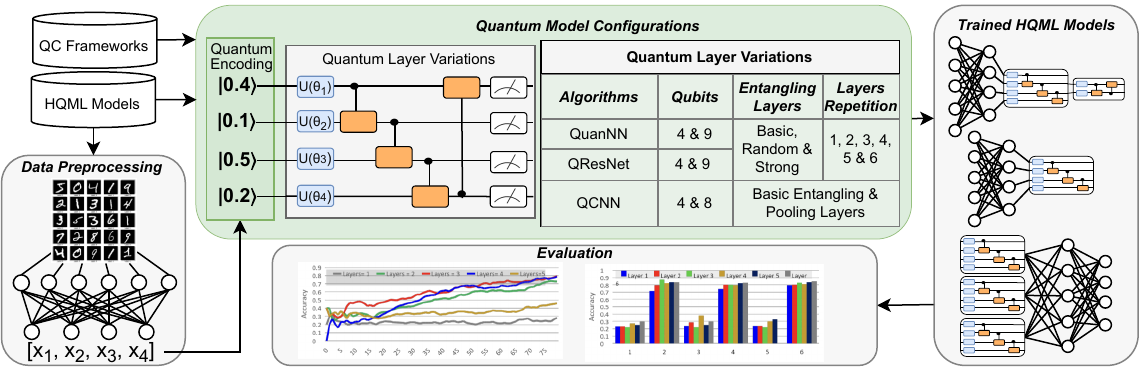}
    \caption{Overview of our comparative analysis methodology.}
    \label{fig:methodology}
\end{figure*}

\subsection{Quantum Convolutional Neural Networks}%~\cite{Cong_2019}}
 Similar to the QuanNN, the Quantum Convolutional Neural Network (QCNN) is a QML algorithm inspired by CNNs that was introduced in~\cite{Cong_2019}. Unlike the QuanNN, the QCNN does not have room for lots of circuit design variations. The QCNN has a fully quantum implementation of convolutional and pooling layers. We chose this architecture in our experiments because it is one of the state-of-the-art QML models and the classical counterpart represents the state-of-the-art in classical image recognition. Moreover, it is important to note that the QCNN presented in~\cite{Cong_2019} makes use of only $O(log(N))$ variational parameters for input sizes of $N$ qubits. This allows for its efficient training and implementation on realistic, near-term quantum devices. 

 The basic structure of the QCNN includes an input encoding circuit layer, a convolutional circuit layer, a pooling circuit layer, and a circuit measurement layer, each composed of parametric quantum gates. It is categorized as an HQML algorithm because it uses classical optimization techniques to update the parameterized gate weights. In the quantum convolutional and pooling layers, the interactions between quantum bits can effectively extract features from the input data based on the types of gates and their placement in each layer. Given the current NISQ devices, the QCNN is limited in terms of scalability and can only be implemented with a small number of qubits. Therefore, it requires classical layers for downsizing large inputs to match the qubit size. In our implementation, we employ single classical convolution and pooling layers for input downsizing, without loss of key features. 

\subsection{Quantum ResNet}%~\cite{liang2021hybrid}}
Our decision to employ the Quantum ResNet (QResNet) algorithm was inspired by~\cite{liang2021hybrid}, which introduced a hybrid quantum-classical strategy for deep residual learning. The primary challenge in this approach is to establish a connection between the residual block structure and the quantum processing layers. In our specific context, our focus lies in experimenting with various quantum layer types. We delve into an analysis of potential methods and variations that combine residual learning with quantum machine learning. Notably, the work in~\cite{Ghasemian2022} emphasizes that experimental outcomes demonstrate the QResNet's superior capability of learning an unknown unitary transformation and its enhanced robustness with noisy data, compared to state-of-the-art methods. These findings motivate us to further explore and fine-tune this promising architecture. Our choice of the QResNet stems from our desire to test this HQML architecture and its impact on pre-trained classical models. 

\section{Comparative Analysis Methodology}
\label{sec:methodology}

With the aim of understanding hybrid-quantum classical neural networks in-depth, we divide our problem into interdependent experimental sections as shown in Fig.~\ref{fig:methodology}. The sectioning allows us to identify areas of possible efficient implementation and improvements for the HQML algorithms discussed in Section \ref{sec:algorithms}.

\subsection{Overview of Analyses}

Our experimental sections are summarized in the \textit{Quantum Layer Variations} table in the \textit{Quantum Model Configurations} group of Fig.~\ref{fig:methodology}. For each algorithm, we analyze the impact of different architectural permutations based on: 
\begin{itemize}[leftmargin=*]
    \item \textbf{\textit{Entanglement variation of quantum circuit}}: The predefined circuit structure of QuanNN and QResNet enables us to change the entanglement type of the circuit. Each circuit has its own orientation of CNOT gates and parameterized corresponding to the strength of their entanglement. Whereas for QCNN, there is no room for changing the entanglement type of the circuit, since it follows the structure defined in \cite{Cong_2019}. 
     
    \item \textbf{\textit{Layer count variation}}: For each algorithm, a circuit can be applied multiple times to an input, which corresponds to understanding how the varying depth of a quantum circuit affects the model accuracy.  
    
    \item \textbf{\textit{Qubit count variation}}: The strength and capability of a circuit depends on the number of qubits it has. Hence, in our experiments, we vary the qubit counts in the architecture to analyze how the variation correlates with the models' learning curve and accuracy. 
\end{itemize}

\subsection{Comparison Metrics} 
 We intentionally made our experimental setup simple and precise, to gain more understanding about how different independent parameters of a circuit's design of our selection algorithms can influence their accuracy. To understand the contribution and convergence behaviours of the aforementioned variations, we focus on analyzing the correlation between a model's accuracy by studying their classical accuracies in relation to the learning curve attained over the training progress. 
 
 %This kind of analysis holds great importance since the quantum machine learning domain is in its very early developmental stages. Hence, at this point, our work is motivated by understanding the underlying behavior and effects of the quantum control parameters to be able to comprehend what factors are driving these hybrid models. Additionally, being in the NISQ era showing avenues of practical potential development direction for HQML is important than setting new state-of-the-art works. %

\section{Evaluation and Discussion}

\subsection{Experimental Setup}
%The scope of this paper defines/limits design variations for the quantum components of the architectures only. 

This paper investigates the design variations of the quantum components of HQML architectures. Hence, to ensure a fair comparison, we use a uniform classical optimization environment for both the classical and the quantum layers as specified in \Cref{tab:exp_setup}. In the HQML architectures of our experiment, all the algorithms have a classical layer after the quantum layer to convert the quantum measurement values into classical probabilities. However, the QCNN has a classical layer before the input encoding for image downsize. As for the quantum layers, PennyLane and Qiskit frameworks provide PyTorch integration modules, which convert a quantum layer into PyTorch trainable layers and perform classical optimization of the gates based on the training hyperparameters that we specified in \Cref{tab:exp_setup}. In our experiments, to ensure that the output of a circuit corresponds to the minimum number of qubits of our qubit count pool, we use a subset of the MNIST dataset \cite{Deng2012TheMD} consisting of only 4 classes, [0, 1, 2, 3]. Thus, the last classical layer of our models consists of only 4 neurons.

For analyzing the HQML components of our architecture, three different kinds of variation are applied to the quantum layer of an algorithm, as shown in Fig.~\ref{fig:methodology}:
\begin{itemize}[leftmargin=*]
    \item \textbf{\textit{Entanglement variation of quantum circuit}}: For algorithms that allow different entanglement orientation, we vary their architectures with Random Circuit (RC), Basic Entangling (BE), or Strongly Entangling (SE) circuit. 
    \item \textbf{\textit{Layer count variation}}: Each quantum layer can have numerous repetitions of a circuit. In our implementation, we repeat the circuit from 1 to 6 times. \textit{Note: the QCNN has a fewer layer variation compared to the QuanNN and QResNet because number of $layer = \sqrt{qubits}$}.
    \item \textbf{\textit{Qubit count variation}}: QuanNN and QResNet circuits follow a square filter-like structure where the qubits must be a square number. Hence, we experiment with 4 and 9 qubits. On the other hand, the circuit of the QCNN must be an even number that can be reduced in half at each pooling step. Hence, we experiment with 4 and 8 qubits.  
\end{itemize}
  
\textit{Note: We implemented all these pipelines in Qiskit~\cite{Qiskit} \& PennyLane~\cite{bergholm2022pennylane} to enable the support for both QML frameworks. They are purposely presented separately (and can be identified in boldened figure captions) to avoid cross-tool comparison which is out of the scope of this paper.}

\begin{table}[!t]
\centering
\caption{Training Environment Specifications}
\label{tab:exp_setup}
\begin{adjustbox}{max width=\linewidth}
\begin{tabular}{c|c}
\textbf{Algorithm} & \textbf{Experiment Name} \\ \toprule
Software FrameWork & \verb|PennyLane(PL)|, \verb|Qiskit(QK)| \\ \midrule
Back-End Simulator & \verb| lightning.qubit(PL)|, \verb|qasm_simulator(QK)|  \\ \midrule
Back-End Machine & \verb| NVIDIA RTX 6000 Ada |\\ \midrule
Deep-Learning Interface & \verb|Pytorch| \\ \midrule
Data-set & \verb|MNIST|~\cite{Deng2012TheMD} \\ \midrule
Training Samples, Testing Samples &  PL: (\verb|100|, \verb|100|), QK: (\verb|500|, \verb|100|) \\ \midrule
Epoch, Batch-Size, LR & \verb|5|, \verb|5|, \verb|0.01| \\ \midrule
\end{tabular}%
\end{adjustbox}
\end{table}

\subsection{Results: Entangling Circuit Variations}
\label{sec:entangling_eircuit_variations}

\begin{figure}[!t]
\centering
\includegraphics[width =\linewidth]{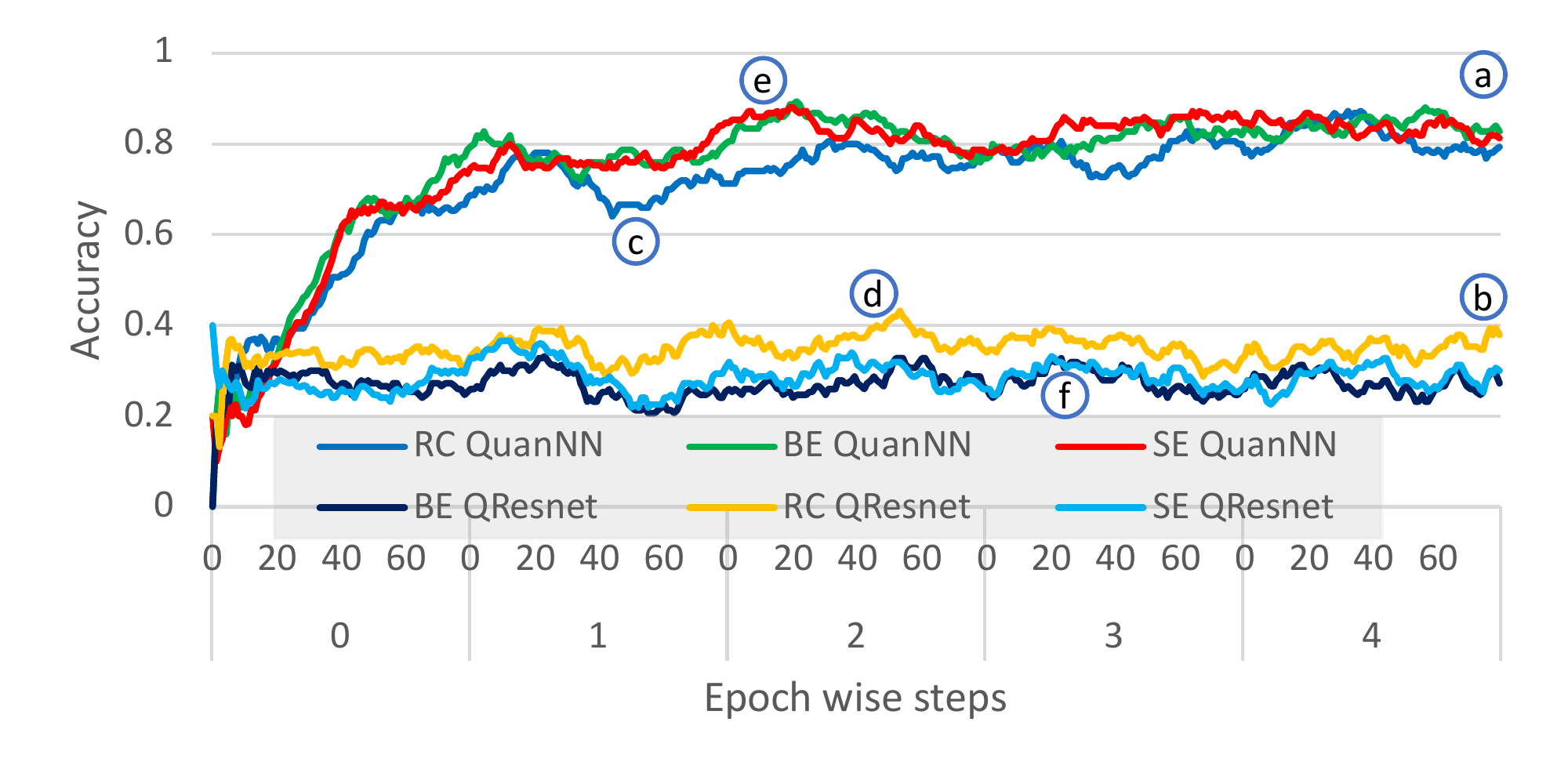}
\caption{\textit{PennyLane:} RC: Random Circuit, SE: Strongly Entangling, BE: Basic Entangling. Circuit Variation across QuanNN and QResNet.} 
\label{fig:circuit_variations}
\end{figure}

Fig.~\ref{fig:circuit_variations} shows the accuracy evolution over training epochs for QuanNN and QResNet at different entanglement settings (RC, BE \& SE) for PennyLane implementation. As demonstrated by the accuracy and convergence gap between the two algorithms, it is evident that QuanNN learns significantly better than QResNet (see labels a and b). 

Further analyses show that the RC QuanNN (label c) acquires lower accuracy than the other entanglement settings. Whereas, for QResNet, the RC model (label d) achieves the highest accuracy compared to the SE and BE settings. Towards the end of the training, we can see that the accuracy of the QuanNN, for all three entanglements, converges to around 80\%, with the SE QuanNN having the highest accuracy. Following the traces for BE and SE circuits, for both QuanNN and QResNet (labels e and f, respectively), it is noticeable how the traces are similar in each algorithm. On the other hand, RC varies in learning compared to SE and BE in both algorithms. In the RC QuanNN (label c), the accuracy improves at a slower pace compared to SE and BE, but soon converges closer towards the end. As for the RC QResNet, the rate of learning is similar to BE and SE, but with a higher accuracy throughout. 

\begin{figure}[!t]
\centering
\includegraphics[width =\linewidth]{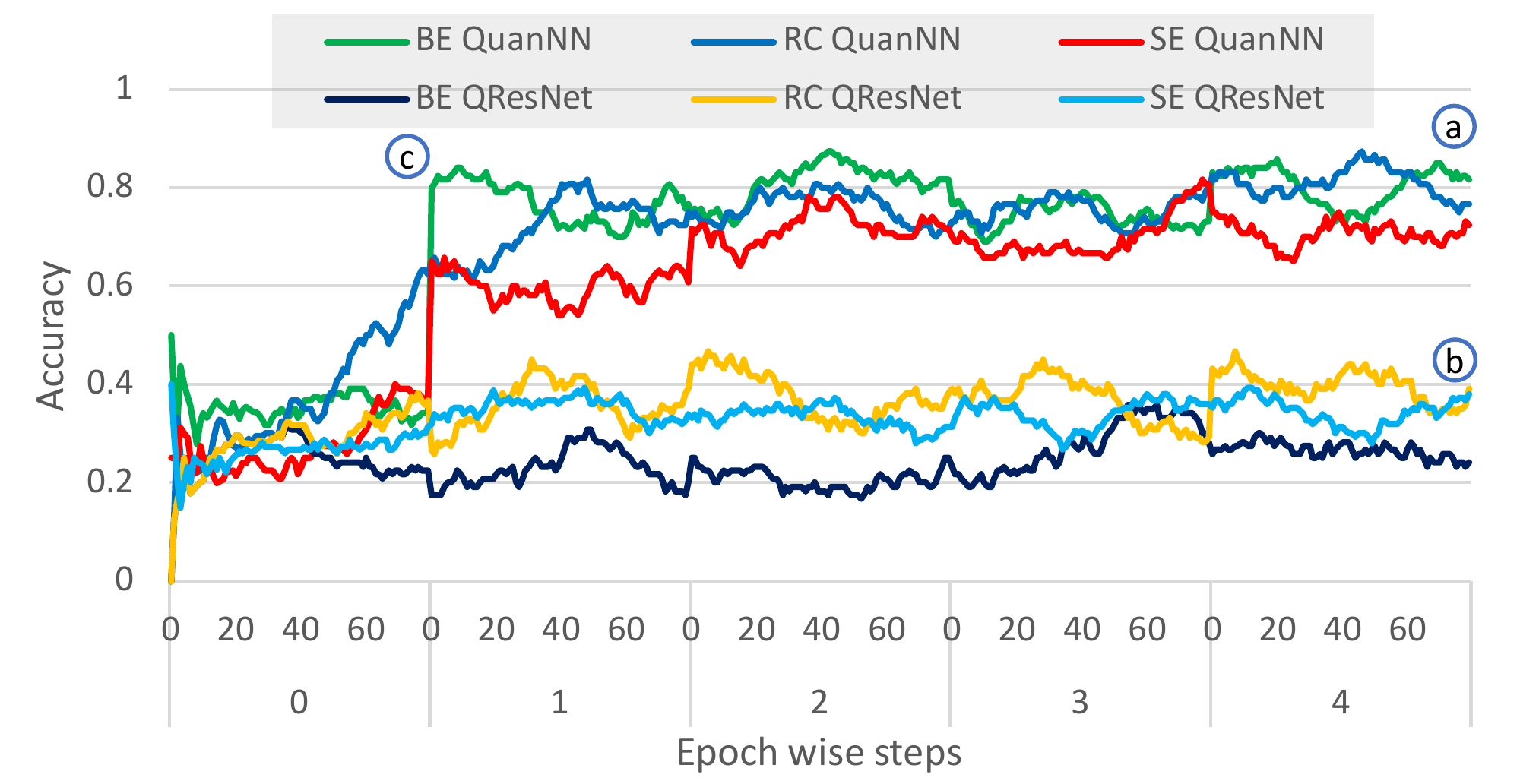}
\caption{\textit{Qiskit:} RC: Random Circuit, SE: Strongly Entangling, BE: Basic Entangling. Circuit Variation across QuanNN and QResNet.} 
\label{fig:qkcircuit_variations}
\end{figure}

Fig.~\ref{fig:qkcircuit_variations} shows the same evolution graphs as Fig.~\ref{fig:circuit_variations}, but for Qiskit models. It is obvious that even for Qiskit implementations the accuracy gap between QuanNN and QResNet models are same as PennyLane (labels a and b). It is quite evident, that QuanNN models in Qiskit have different learning curve, with a great accuracy jump after the first epoch (label c).

\textit{Note: the QCNN is not presented in this comparison because only the Basic Entangling circuit can be implemented in this architecture.}

\subsection{Results: Layer Count Variations}
\label{sec:layer_count_variations}

\begin{figure}[!t]
\centering
\includegraphics[width =\linewidth]{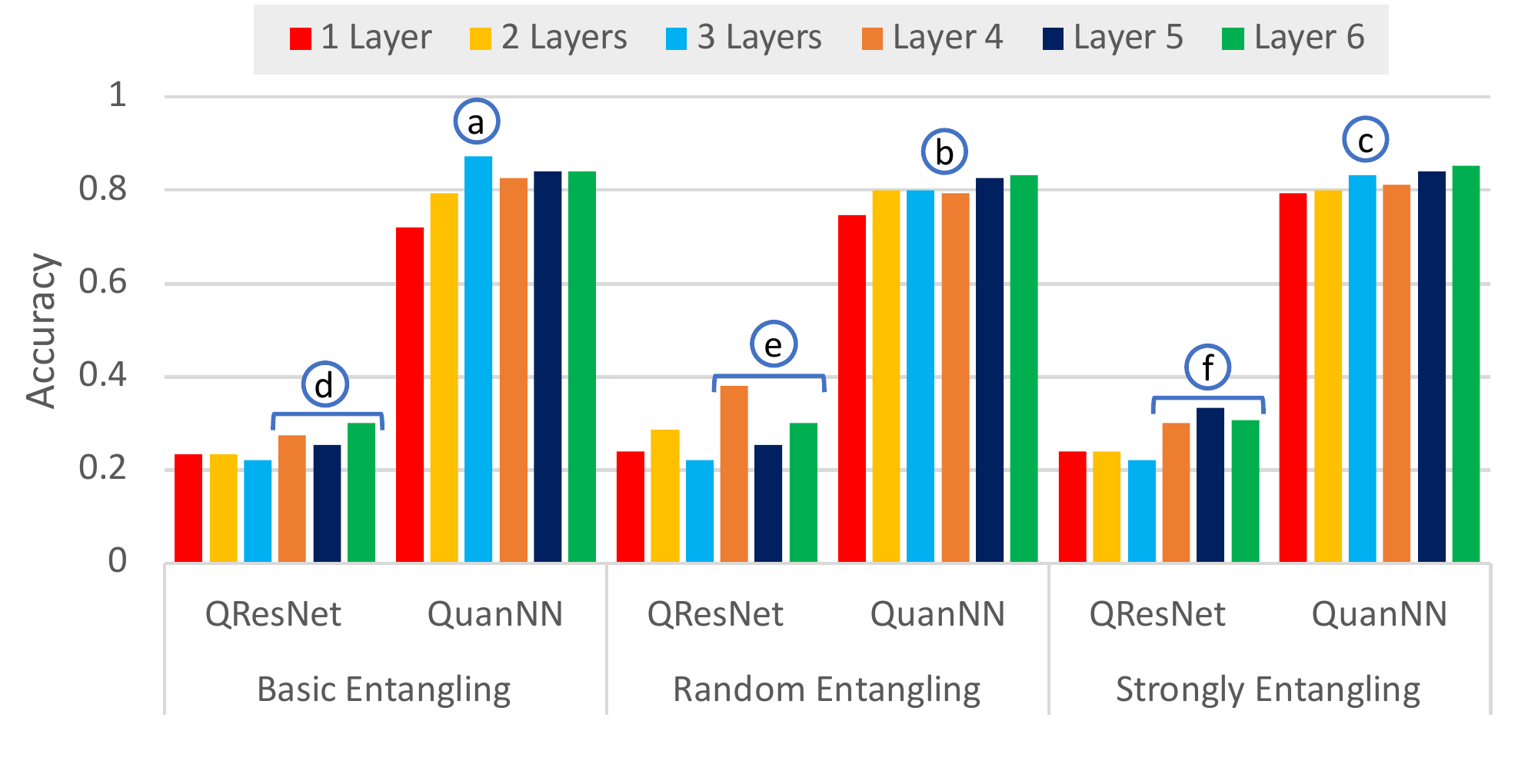}
\caption{\textit{PennyLane:} Layer Variation across different Entangling Circuits for QuanNN and QResNet. Asides minor perturbation, an overall trend of improving accuracy is observed with the increasing number of layers, for every type of circuit.}
\label{fig:pl_layercount}
\end{figure}

Fig.~\ref{fig:pl_layercount} shows the accuracy of the QResNet and QuanNN models with varying number of layers implemented in PennyLane. From the results, we can gauge that there is minor difference between 2, 3, and 4 layers for QResNet models (see labels b and c). Moreover, adding more than 4 layers to the QuanNN does not always contribute to increasing the accuracy (see labels a and e). For the QResNet, it can be observed that the accuracy increases quite significantly for the QResNet models with 4, 5, and 6 layers (see d, e, and f), more specifically for circuit with 4 layers. Based on this observation, we conducted the remaining experiments with 4 layers in places where no layer variation was involved. % On the other hand, using only 1 layer does not guarantee acceptable accuracy. Hence, we do not envision useful usage of 1-layer networks. 
Even though the accuracy of the QResNet is lower than the QuanNN, the algorithm has some potential because it can improve the accuracy for increasing the number of layers (see labels~d and~f).

\begin{figure}[!t]
\centering
\includegraphics[width =\linewidth]{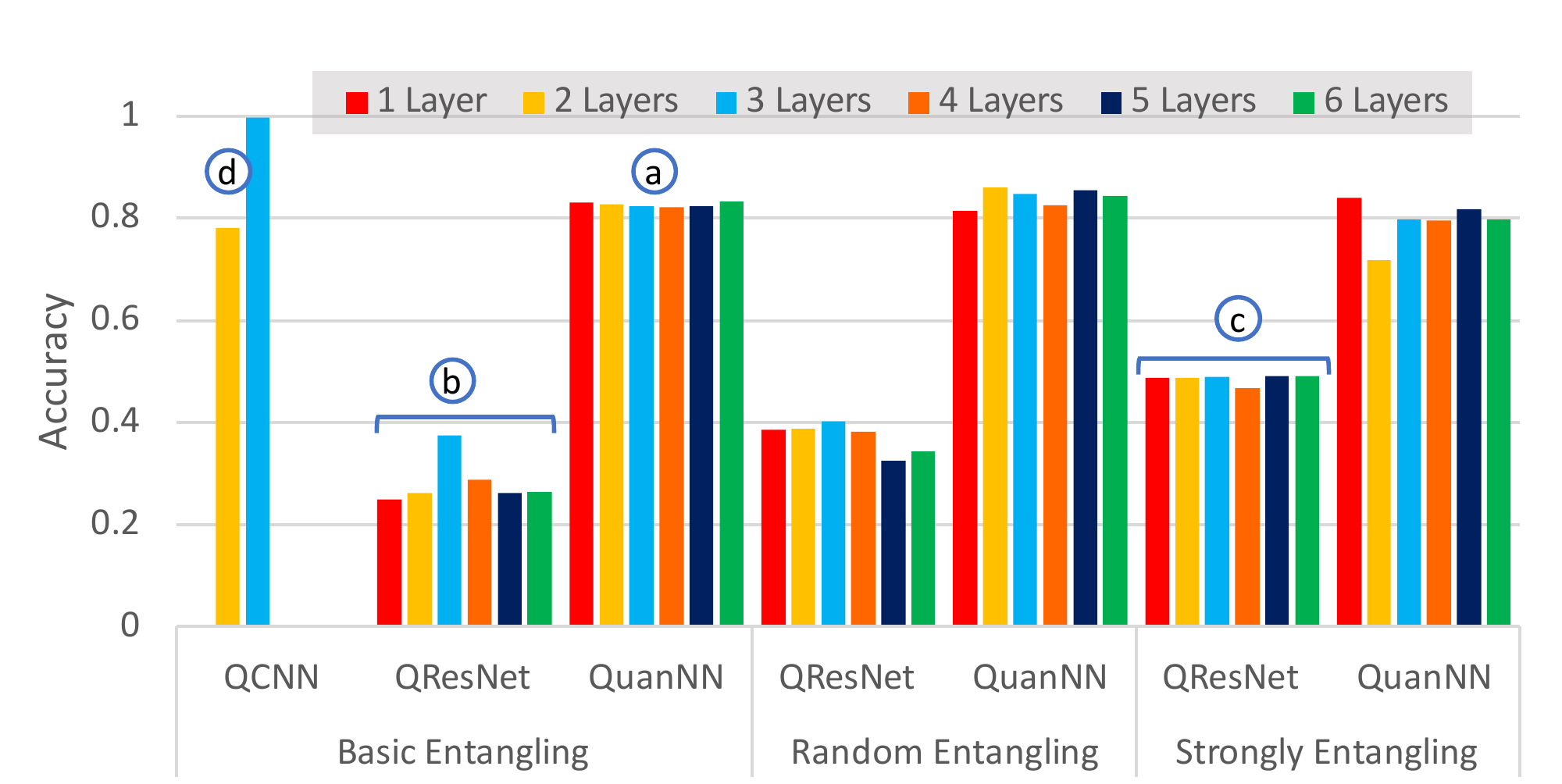}
\caption{\textit{Qiskit:} Layer Variation across different Entangling Circuits for QuanNN and QResNet. No overall trend observed with increasing number of layers.}
\label{fig:qiskit_variations}
\end{figure}

In Fig.~\ref{fig:qiskit_variations}, we can observe that the QuanNN and QResNet models implemented in Qiskit do not have an overall trend with respect to increasing the number of layers, which indicates that the circuit depth does not have enough impact on these models' accuracy (see labels a, b and c). However, it is quite evident for QCNN models that increasing the number of layers has a positive correlation with the accuracy (see label~d). %\textit{Note: the QCNN has a fewer layer variation compared to the QuanNN and QResNet because number of $layer = \sqrt{qubits}$}

\begin{figure}[!t]
\centering
\includegraphics[width =\linewidth]{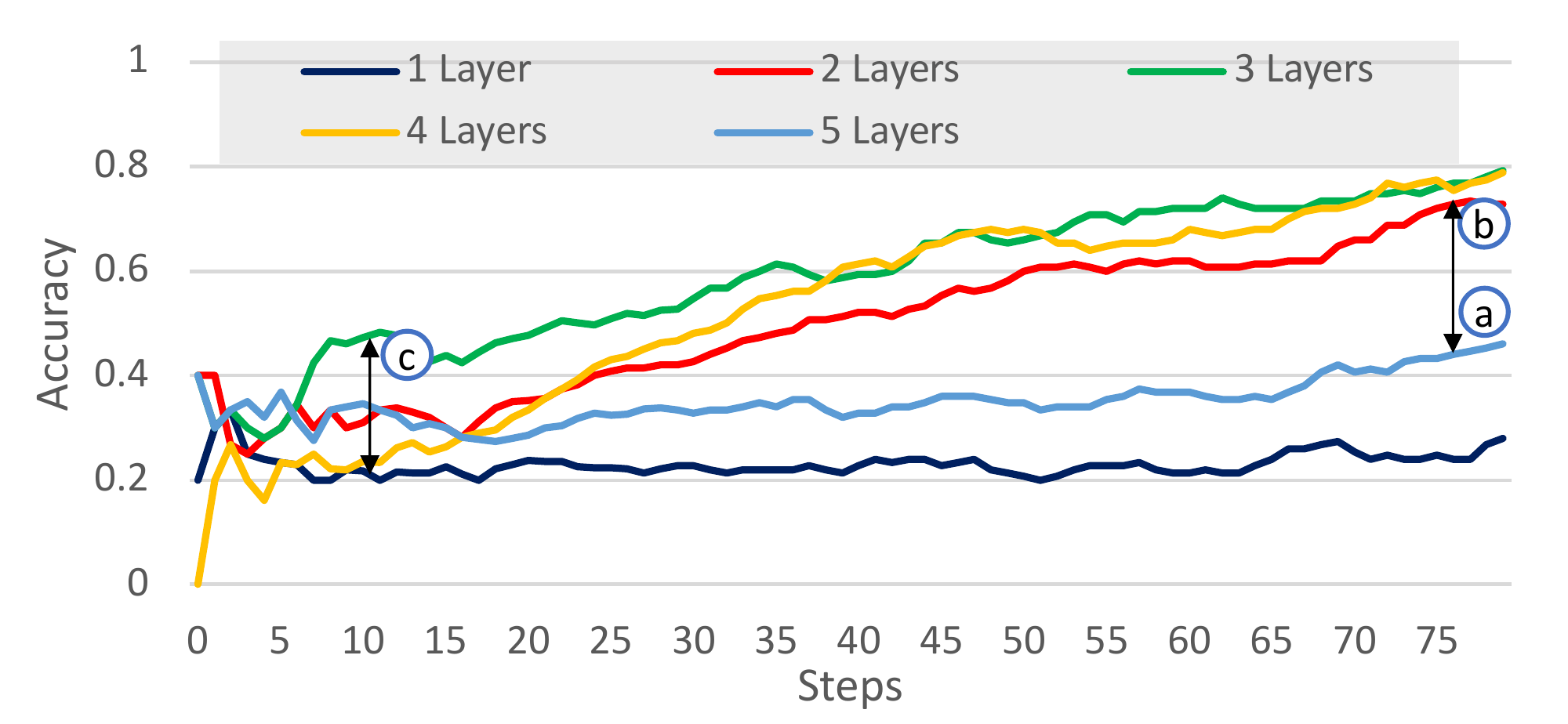}
\caption{\textit{PennyLane:} Initial learning behavior of Layer Variations, for the QuanNN with Basic Entangling. Having 3 layers shows a clear advantage over the other configurations in the first epochs, but the accuracy of the 4-layer QuanNN, despite being slow at first, catches up quickly with the 3-layer counterpart and converges to better results in the majority of the experiments.}
\label{fig4_penny}
\end{figure}

Fig.~\ref{fig4_penny} shows the learning curve for the QuanNN with Basic Entangling for different layers implemented in PennyLane. For 1 layer, the learning curve is constant throughout and little improvement is observed (see label c). For more than one layer, we can observe that the learning improves with increasing the number of layers. However, the optimal peak is until 4 layers (see label b), because beyond that the learning curve drops to being close to the 1-layer curve (see label a). Nonetheless, a steady and gradual improvement in accuracy is observed for more than one layer. 

%\begin{itemize}
    %\item PennyLane observations and chart
    %\item  Basic Entangling optimal up to 4 layers then starts dropping again
    %\item for Random and Strongly Entangling circuits improvement trend with increasing layers 
    %\item Layer increase shows haphazard unpredictable behavior for ResNet some underlying trend is always present for QuanNN.  
%\end{itemize}

\begin{figure}[!t]
\centering
\includegraphics[width =\linewidth]{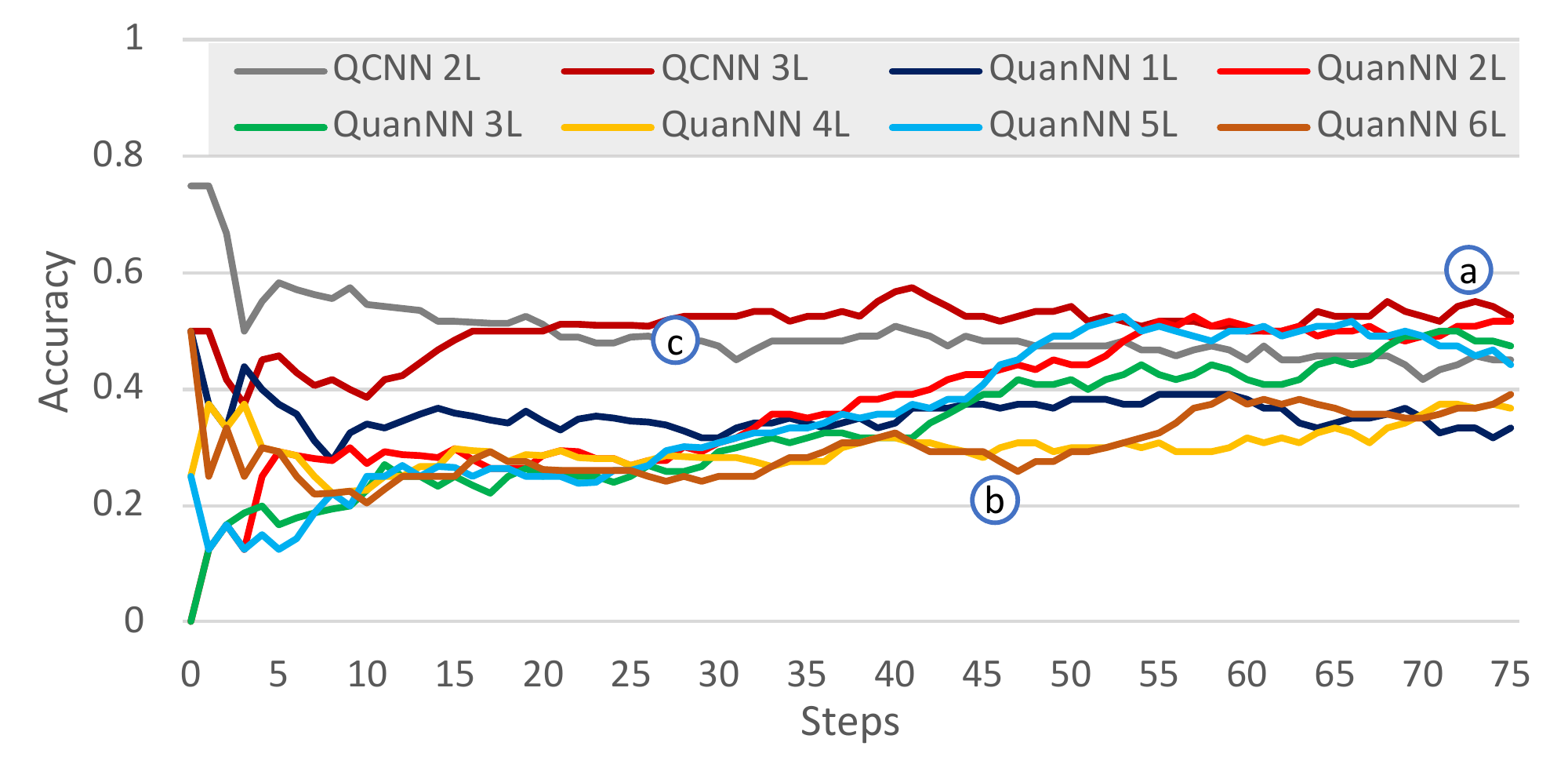}
\caption{\textbf{Qiskit:} Initial learning behaviour of different Basic Entangling layer variation counts.}
\label{fig4_qiskit}
\end{figure}

For the models implemented in Qiskit, as shown in Fig.~\ref{fig4_qiskit}, we can observe that QCNN models with 3 layers have the highest accuracy towards the end of the first epoch (see label a), however, with a very steadily increasing learning curve (see label c). As for the QuanNN models, although the initial accuracy is lower than the QCNN 3L, the models have a rapidly increasing learning curve. Moreover, we can see that the learning curve for the QuanNN models peak only until 5 layers (QuanNN 5L), because beyond that, for 6 layers, a very slow and steady improvement is observed (see label b). 

\subsection{Results: Qubit Count Variations}
\label{sec:qubit_count_variations}

\begin{figure}[!t]
\centering
\includegraphics[width =\linewidth]{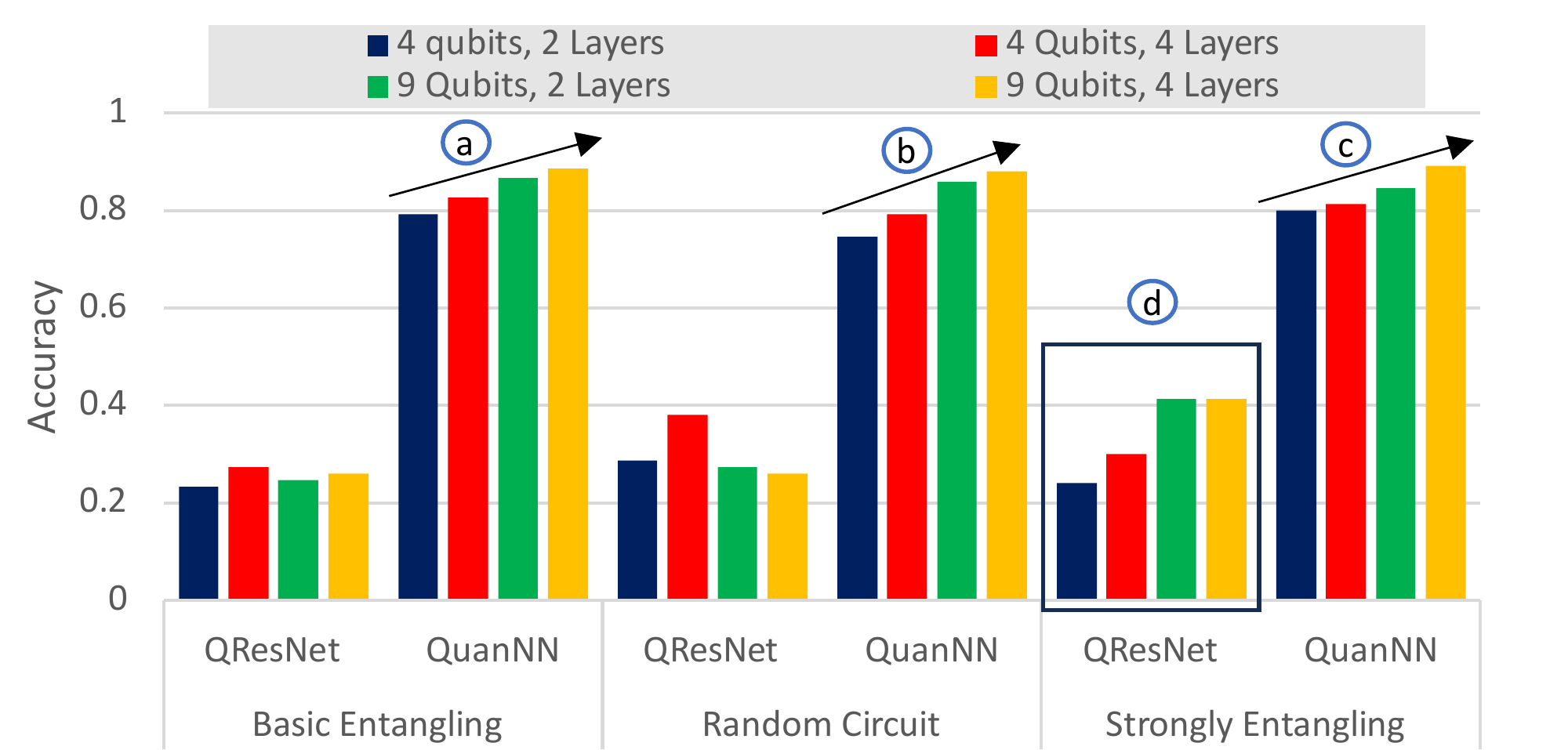}
\caption{\textit{PennyLane:} Qubits Count Variation across different quantum entangling circuits for QuanNN and QResNet. It can be noted that there a positive correlation between number of qubits in QuanNN model and its accuracy.}
\label{fig:qubit_variations}
\end{figure}

Fig.~\ref{fig:qubit_variations} illustrates the results for the QuanNN and QResNet models with different qubits counts, implemented in PennyLane. A clear pattern can be observed for the QuanNN models, where there is a consistent positive correlation between the number of qubits and model accuracy, with increasing layer counts (see labels a, b, and c), indicating that the models learn better with increasing number of qubits. On the other hand, the QResNet models with Basic Entangling circuit and Random Circuit do not have any significant trend. However, the QResNet with Strongly Entangling circuit shows an increase in accuracy with increasing its qubit count (see label d), which might lead one to derive that QResNet with Strongly Entangled circuit has a positive learning capability with an increased number of qubits. 

%\item We tested qubits 4, 9 and 16, where 4 and 9 have been easily implemented but their feasibility test run for 16 qubits on 20 samples although successful in completion failed due to extensively long run time required. It is a compute-intensive task to simulate 16 qubit quantum circuits for deep learning layers over several images.

\begin{figure}[!t]
\centering
\includegraphics[width =\linewidth]{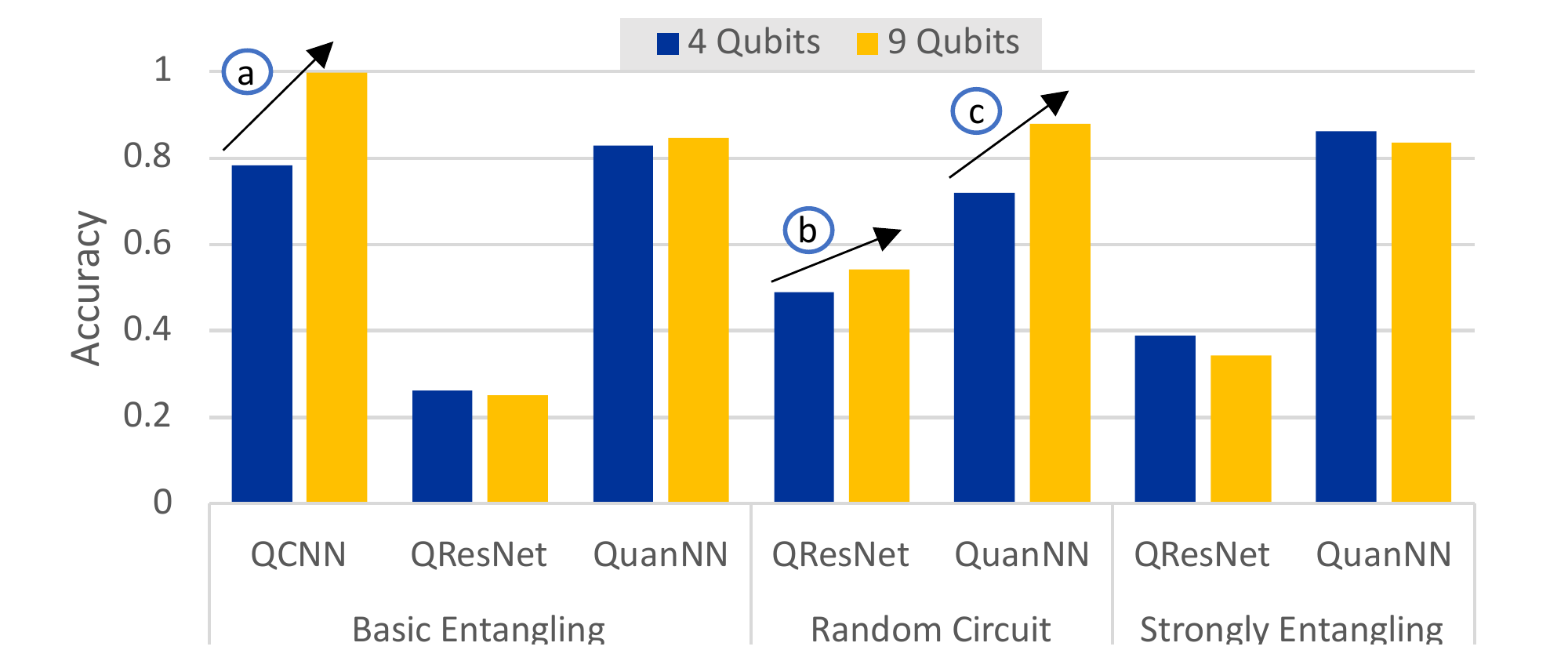}
\caption{\textit{Qiskit:} Qubit Count variations across different entangling circuits for QuanNN, QResNet, and QCNN. The QCNN accuracy positively correlates with the increasing number of qubits.}
\label{fig:qcn_qubit}
\end{figure}

Fig.~\ref{fig:qcn_qubit} depicts the variation of models implemented in Qiskit with regards to varying qubit count. Firstly, we can notice that the QCNN accuracy improves with increasing the number of qubits, which indicates that the QCNN learns better with more qubits (see label a). A similar trend is observed for the QuanNN and QResNet, but only for the models with Random Circuit (see labels b and c). From this analysis, we can gauge that, changing the number of qubits in Qiskit models have a varying impact for different entangling circuits.

\subsection{Result Discussion \& Findings}

We observed from the results that different variations combined together have a compounding effect on the model accuracy, as it is not trivial to determine the best set of permutations for the given HQML algorithm. For instance,  the QResNet had a stagnant improvement across most models. On the other hand, if implemented with Strongly Entangling and 9 qubits, then the accuracy improves drastically. However, in most cases, the accuracy gain achieved when increasing the number of qubits comes with the cost of increased execution time.

In addition, varying the number of layers with different entangling circuits can impact the accuracy differently, as we observed in our results above. In some cases, the increasing accuracy with increasing depth indicates that there is potential in applying more quantum layers at the initial stages of HQML models instead of classical convolutional layers. However, as observed in Fig.~\ref{fig:qubit_variations}, we can see that the accuracy of HQML models keeps increasing as the model becomes more complex (i.e., with a higher number of layers and qubits). However, the more complex a model becomes, the more time it takes to execute. 

All of our findings suggest that the best circuit configuration cannot be identified easily for a given algorithm when keeping in mind all the constraints. %With the results of our analysis, we aim to pave a pathway for the development of a more robust Hybrid Quantum-Classical algorithm by analyzing different circuit configuration and their scope for improvement in learning capabilities.

 %Moreover, in some cases, the similarity in behavior of Basic Entangling and Strongly Entangling circuits make them redundant to one another. In such cases, it is sufficient to merge these two configurations due to the minor differences between each other.

%Despite that, it seems evident that Basic Entangling and Strongly Entangling circuits are redundant to each other and can be used interchangeably. The underlying system they operate on is resulting in a very closely similar behavior. Therefore, it is sufficient to merge these two configurations due to the minor differences between each other. These types of simplifications are useful to reduce the granularity of the hybrid quantum circuits and provide useful insights. Such an in-depth analysis serves as the building block for designing robust and targeted HQML architectures. %within the context of the image classification problem. 
%Hence, to maximize the strong suits of each parameter and structure along with handling the trade-offs reasonably without affecting accuracy providing a sound base for designing novel hqml architectures with understanding of choice of parameters defining those architectures.

\section{Conclusion}

In our work, we studied the accuracy variation of 3 different algorithms, QuanNN, QCNN, and QResNet, by configuring the quantum architecture of each circuit by varying the layer count, qubit count, and the type of entangling circuit. These 3 categories of variation enabled us to design different permutations of circuits for a given algorithm and study their effect on their accuracy.

This study has granted us valuable insights into the establishment and exploration of avenues for optimizing hybrid quantum-classical neural network architectures. We have effectively advanced our pipelines toward practical improvements, using foundational structures for experimentation. The detailed deconstruction of the architecture offers transparency into the parameters that influence the performance of quantum circuits within neural networks. Such an in-depth analysis serves as the building block for designing efficient HQML architectures and paves the way to develop robust HQML algorithms applicable in the NISQ era, while also considering design aspects that can smoothly transition into Fault Tolerant Quantum Computers.

\section*{Acknowledgments}

This work was supported in part by the NYUAD Center for Quantum and Topological Systems (CQTS), funded by Tamkeen under the NYUAD Research Institute grant CG008.

\bibliographystyle{ieeetr}
\bibliography{main.bib}

\end{document}